\documentclass[]{iopart}
\usepackage{graphicx}
\usepackage{color}

\begin{document}

\title{Doppler-free two-photon resonances for atom detection and sum frequency stabilization}

\author{A. M. Akulshin, B. V. Hall, V. Ivannikov, A. A. Orel, and A. I. Sidorov}

\address{Centre for Atom Optics and Ultrafast Spectroscopy,\\
Swinburne University of Technology, Hawthorn, Victoria 3122,
Australia}

\ead{aakoulchine@.swin.edu.au}

\begin{abstract}

We investigate the excitation of the 5$D_{5/2}$ level in Rb atoms using counter-propagating laser beams, which are nearly resonant to the one-photon   $5S_{1/2}\rightarrow5P_{3/2}$ and $5P_{3/2}\rightarrow5D_{5/2}$ transitions, ensuring that a sum of the optical frequencies  corresponds to the $5S_{1/2}\rightarrow5D_{5/2}$ transition. The  excitation produced by two-photon and step-wise processes is detected via spontaneously emitted fluorescence at 420 nm arising from the $6P_{3/2}\rightarrow5S_{1/2}$ transition.
The dependences of blue fluorescence intensity on atomic density and laser detuning from the intermediate $5P_{3/2}$ level have been investigated. The sensitivity of the frequency detuned bi-chromatic scheme for atom detection has been estimated. A novel method for sum frequency stabilization of two free-running lasers has been suggested and implemented using two-photon Doppler-free fluorescence and polarization resonances.

\end{abstract}

\pacno{39.90.+d,03.75.Be,39.25.+k,07.55.-w}
\date{\today}

\maketitle 

\section{Introduction}

Optical detection of a few atoms relies predominantly on signals generated by spontaneously emitted photons \cite{single}.  While a resonant laser radiation facilitates optical excitation, it can also readily produce a large background due to scattering of light from optical elements. If the weak signal and the unwanted scattered light are not spectrally separated then heroic efforts must be undertaken to obtain a reasonable signal-to-noise ratio while ensuring the high collection efficiency of spontaneously emitted photons that is crucial for quantum information experiments \cite{efficiency}. This may become unrealistic for fluorescence imaging of single atoms located near surfaces on atom chips or in the microscopic pyramidal magneto-optical traps \cite{Pol09}.

\begin{figure}
\begin{center}
\includegraphics[angle=0, width=9cm]{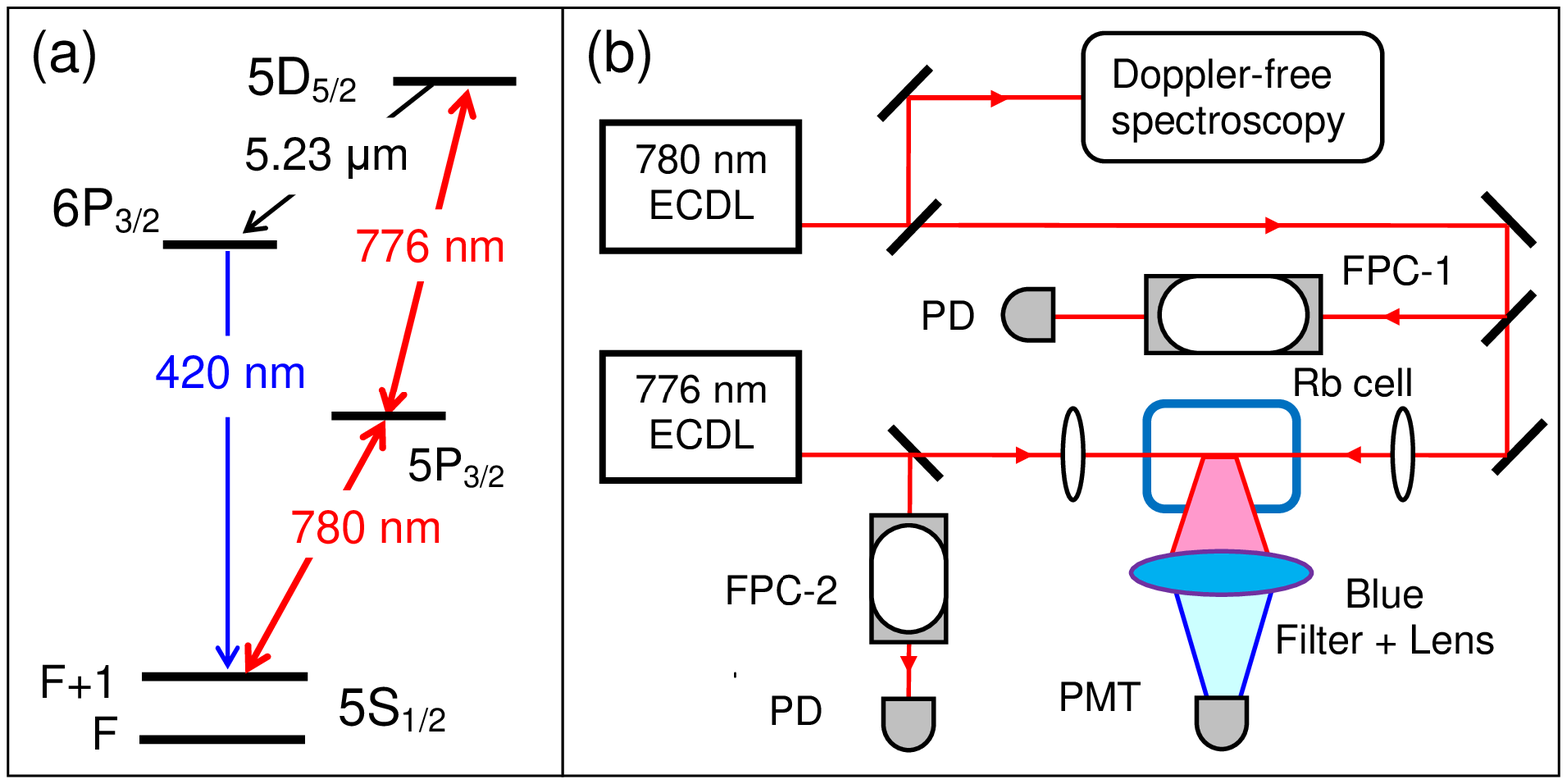} \end{center}
\caption{a) Energy level diagram for the optical transitions in Rb atoms used in the experiment. b) Schematic of the spectroscopy setup. \label{Figure1}}
\end{figure}

An alternative approach is to utilize two-photon excitation to access higher laying atomic states which decay by emitting photons with sufficiently different frequency. For example, in rubidium atoms the two-photon transition $5S_{1/2} \rightarrow 5D$ (Fig.~\ref{Figure1}a) can be realized using laser radiation at 778~nm, while subsequent decay via the $6P_{3/2}$ and $6P_{1/2}$ levels yields spontaneously emitted light at 420~nm, which can be effectively separated from the applied laser radiation using color or interference filters. The intermediate $5P_{3/2}$ energy level, situated nearly half-way between the $5S_{1/2}$ and $5D$ levels, makes this two-photon transition strongest amongst alkali atoms.

The two-photon transition $5S_{1/2}\rightarrow 5D_{5/2}$ in Rb atoms \cite{Nez93} has found a wide range of applications.  Due to a relatively long natural lifetime of the $5D_{5/2}$ level ($\sim$240 ns) the two-photon transition $5S_{1/2}(F=3)\rightarrow5D_{5/2}(F"=5)$ in $^{85}$Rb, where $F$ and $F"$ are the total angular momenta, was recommended as a secondary frequency reference \cite{Edw05}. Also this transition was suggested to be used for transferring long-term frequency stability to the telecommunication spectral region at 1.5 $\mu$m \cite{telecom}. The two-photon transition was employed for studying the excitation transfer between the Rb $5D$ fine-structure levels in collisions with ground-state atoms \cite{Bieniak}. This scheme was also used to populate the $5D$ level in a magnetic trap whereby photoionization and subsequent ion detection yielded atom counting capability \cite{Kra07}. The two-photon excitation to the 5$D_{5/2}$ level was probed, in addition to blue fluorescence, by the polarization rotation of the laser light due to the induced optical anisotropy of the medium \cite{Ramiz03}.

More efficient transfer of atoms from the $5S_{1/2}$ to $5D_{5/2}$ levels and, consequently, more intense blue fluorescence can be obtained using a bi-chromatic excitation scheme, which employs radiation resonant to the one-photon transitions $5S_{1/2}\rightarrow 5P_{3/2}$ and 5$P_{3/2}\rightarrow 5D_{5/2}$, ensuring that the sum frequency of the two laser fields ($\nu_1$ + $\nu_2$) equals to the two-photon transition frequency. In this case, in addition to the enhanced two-photon excitation rate due to proximity of the intermediate level 5$P_{3/2}$, there is another process, step-wise excitation, which plays an important role when the frequency detunings of the applied radiation from the one-photon transitions are comparable with inhomogeneous broadening of the transition.

The bi-chromatic excitation scheme has received a lot of attention. If the intermediate state 5$P_{3/2}$ is coupled by both components of the bi-chromatic radiation a good signal-to-noise ratio of Doppler-free spectra of the Rb $5D_{5/2}$ level can be obtained even in a room temperature vapour \cite{Gro95}. The bi-chromatic  excitation scheme has been used for in situ imaging an ultracold atom cloud by measuring absorbtion at 776~nm or by observation of the 420~nm fluorescence \cite{She07,Oha09}. Electromagnetically induced transparency (EIT) in cascade atomic system, when the absorption of a probe beam resonant to the lower cascade transition is reduced under the action of a coupling beam resonant to the higher transition, has been studied \cite{Gea95}.

It is worth noting that the co-propagating instead of counter-propagating scheme of the bi-chromatic excitation of the Rb $5D_{5/2}$ level  allows an optical field generation at 420 nm, which is both spatially and temporally coherent \cite{Mei06, Aku09, Ver10}. Taking into account that  detection of the two-photon $5S_{1/2}\rightarrow 5D_{5/2}$ monochromatic excitation based on the stimulated emission at 776 nm was recently demonstrated \cite{Sanguinetti07}, possible applications of coherent light at 420 nm for atom detection deserves a detail consideration. This will form the subject of a subsequent paper. In this paper we study bi-chromatic excitation of rubidium  atoms in a vapour cell by nearly-resonant laser radiation with a subsequent detection of spontaneously emitted blue fluorescence paying a particular attention to the case when the laser frequency detuning is larger than the inhomogeneous broadening of the Rb $D_2$ absorption line.

This paper is structured as follows. In Section 2 a description of the experimental apparatus is given along with the methods for acquiring and comparing step-wise and two-photon processes.  Section 3 presents measurements of the blue fluorescence as a function of laser frequency detunings from single-photon transitions. The fluorescence signal behavior with laser intensity, polarization and atomic density along with detection capability of this method is discussed. In Section 4 a novel sum-frequency locking of two free-running lasers to the $5S_{1/2}\rightarrow 5D_{5/2}$ transition in Rb atoms is considered. Doppler-free fluorescence or dispersive-type polarization resonances are used as frequency references for the sum frequency stabilization.

\section{Excitation scheme and experimental setup}

The two-photon excitation of Rb atoms in a  vapour cell is produced by a bi-chromatic  laser radiation whose components at 780 nm and 776 nm are near-resonant to the one-photon transitions 5$S_{1/2}\rightarrow5P_{3/2}$ and  5$P_{3/2}\rightarrow5D_{5/2}$ respectively. The relevant energy levels and optical transitions are depicted in Fig. \ref{Figure1}a. After excitation to the 5$D_{5/2}$ level, 65$\%$ of Rb atoms spontaneously decay back to the 5$P_{3/2}$ level, while the remaining atoms decay first to the 6$P_{3/2}$ level and then to the ground state emitting blue photons at 420~nm \cite{Hea61}. Counter-propagating geometry of the applied laser beams results in much stronger blue fluorescence because of partial Doppler shift compensation for laser beams with close wavelengths \cite{Bjo76}. Blue fluorescence and a signal proportional to the rotation of the polarization of the 776 nm laser beam passed through the atomic media are used for detecting the two-photon excitation.

The scheme of the apparatus is shown in Fig.~\ref{Figure1}b. Two home-built external cavity diode lasers supply output powers of approximately 20~mW at 780~nm and 776~nm .  The absolute frequency and the scan range of the 780~nm laser is evaluated based on Doppler-free absorption spectra obtained from an ancillary Rb cell and transmission resonances of a tunable low-finesse Fabry-Perot cavity (FPC1). The frequency of the 776~nm laser is monitored with another low-drift Fabry-Perot cavity (FPC2) with the free spectral range of 710 MHz and spectroscopic signals from the $6P_{3/2}\rightarrow5D_{5/2}$ transition.

The excitation occurs in a vapour cell which contains natural mixture of both $^{85}$Rb and $^{87}$Rb isotopes without any buffer gas. The interaction region from which the fluorescence is collected is in the center of the 50~mm long cell. A specially designed heating/cooling thermoelectric element based on the Peltier effect allows varying the cell temperature from 10 to 70 $^0$C. Thus,  atomic density estimated based on the temperature measured in the coldest part of the cell  is in the range ($2\times 10^{9}\leq N \leq 7\times 10^{11}$~cm$^{-3}$) \cite{Rb_Data}.

The counter-propagating laser beams are parallel or focused to waists of approximately 250~$\mu$m. Control of the polarization and intensity of the laser beams is achieved using polarizing beam splitters and wave plates not shown in Fig.\ref{Figure1}b for simplicity. A 420~nm bandpass interference filter and colour filters ensure that only blue fluorescence is detected by the Hamamatsu photomultiplier tube (PMT). Spectroscopic signals  are recorded and processed using a digital oscilloscope and a computer.  Amplitude fluorescence noise uncorrelated to laser frequency is reduced by applying signal averaging over 32 or 64 periods of frequency scans. In order to further increase the signal-to-noise ratio a lock-in amplifier is used. Both amplitude and frequency modulation methods with corresponding modulation frequencies of 700 Hz and 10 kHz are employed.

\section{Results and discussion}

\begin{figure}
\begin{center}
\includegraphics[angle=0, width=9cm]{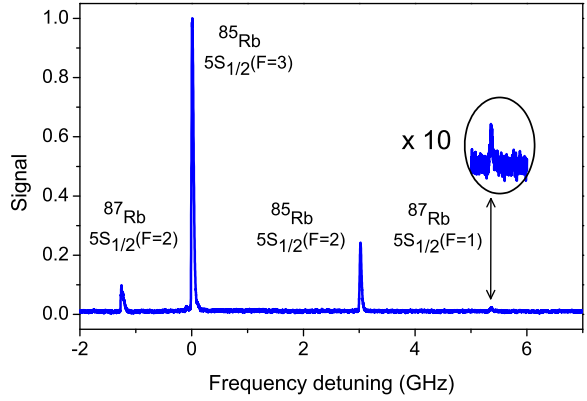}
\end{center}
\caption{The bi-chromatic two-photon excitation of Rb atoms in a vapour cell. Curve represents blue fluorescence signal as a function of the 776 nm  laser frequency scanned in the vicinity of the $5P_{3/2}\rightarrow 5D_{5/2}$ transitions, while the fixed frequency 780 nm laser is approximately 1.2 GHz blue detuned from the 5$S_{1/2}(F=3)\rightarrow 5P_{3/2}(F'=4)$ transition.  \label{Figure2}}
\end{figure}

\subsection{Two-photon resonances on the $5S_{1/2}\rightarrow 5D_{5/2}$ transition}

To demonstrate the whole spectrum of the blue fluorescence arising from the $5S_{1/2}\rightarrow 5D_{5/2}$ transition in Rb vapour under the nearly-resonant bi-chromatic excitation a wide frequency scan of the 776 nm laser is applied, while the fixed frequency 780 nm laser is tuned to the spectral region between the $^{85}$Rb absorption lines on the $5S_{1/2}(F=3)\rightarrow 5P_{3/2}$ and $5S_{1/2}(F=2)\rightarrow 5P_{3/2}$ transitions. Figure \ref{Figure2} shows four resonances of blue fluorescence, which occur when the sum frequency of the two lasers ($\nu_1$ + $\nu_2$) is equal to the frequencies of the transitions started from each ground-state sublevel of the both Rb isotopes to the $5D_{5/2}$ levels. The strongest fluorescence resonance corresponds to the excitation of $^{85}$Rb atoms from the $5S_{1/2}(F=3)$ level. The relative amplitudes of fluorescence resonances depend on the frequency offset of the 780 nm laser from the one-photon transitions within the Rb $D_{2}$ absorption line. This issue is considered in the next subsection. Insert shows the 10-times amplified fluorescence feature attributed to the excitation of $^{87}$Rb atoms from the $5S_{1/2}(F=1)$ ground-state sublevel.

\begin{figure}
\begin{center}
\includegraphics[angle=0, width=6cm]{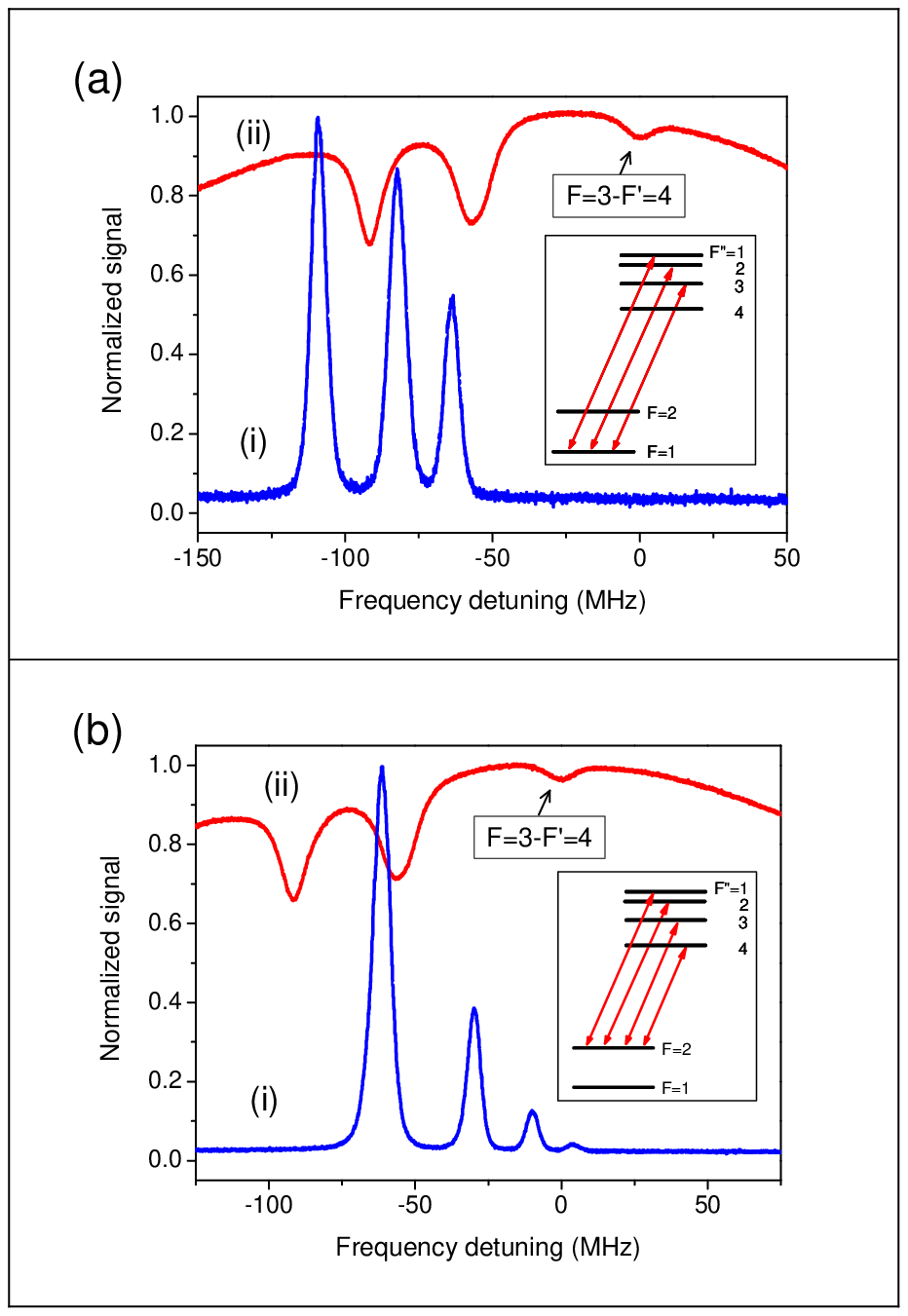}
\end{center}
\caption{Two-photon excitation of $^{87}$Rb atoms from (a) the  $5S_{1/2}(F=1)$ to $5D_{5/2}(F"=1, 2, 3)$  and from (b) $5S_{1/2}(F=2)$ to $5D_{5/2}(F"=1, 2, 3, 4)$ levels. Curves (i) represents the normalized blue fluorescence as a function of the 780 nm laser detuning from the $5S_{1/2}(F=3)\rightarrow 5P_{3/2}(F'=4)$ transitions in $^{85}$Rb.  The saturated absorption resonances (curves (ii)) obtained in the axillary cell serve as frequency references. \label{Figure3}}
\end{figure}

The hyperfine structure of the $5D_{5/2}$ level remains unresolved due to the fast wide-range laser frequency scan, however, with a smaller frequency scan every peak of blue fluorescence presented in Fig. \ref{Figure2} reveals a structure. As an example, Fig. \ref{Figure3} demonstrates blue fluorescence obtained with $^{87}$Rb atoms excited from $5S_{1/2}(F=1)$ and $5S_{1/2}(F=2)$ ground-state sublevels, respectively, as a function of frequency detuning of the 780 nm laser from the $5S_{1/2}(F=3)\rightarrow 5P_{3/2}(F'=4)$ transition in $^{85}$Rb. The spectrum of blue fluorescence shown in Fig. \ref{Figure3}a is obtained when the fixed frequency 776 nm laser is blue detuned by approximately 5.7 GHz from the 5$P_{3/2}\rightarrow 5D_{5/2}$ transition, while Fig. \ref{Figure3}b demonstrates spectral dependence of blue fluorescence with 1.1 GHz red-detuned laser frequency from the same transition.

The selection rule for two-photon transitions ($\Delta F \leq 2$) allows  three transitions from the lower ground-state sublevel $5S_{1/2}(F=1)$ and four transitions from the higher ground-state sublevel $5S_{1/2}(F=2)$ to the $5D_{5/2}(F"=1, 2, 3, 4)$ levels.  Figure \ref{Figure3} demonstrates well resolved hyperfine structure of the  $5^2D_{5/2}$ level. The frequency intervals  between peaks, measured using a frequency scale that is provided by the saturated absorption spectrum on $5S_{1/2}(F=3)\rightarrow 5P_{3/2}$ transitions recorded simultaneously in the axillary cell reference, correspond within 10\% to the frequency splitting of the  $5D_{5/2}$ level \cite{Gro95}. The discrepancy is due to non-linearity of the laser frequency scan.

The width of the fluorescence resonances depends on frequency detuning from the intermediate $5P_{3/2}$ level. The ultimate linewidth of the Doppler-free resonances produced by the monochromatic two-photon   excitation is defined solely by a sum of the initial and final states lifetime $\gamma_S+\gamma_D$, where is $\gamma_D\simeq~2\pi~\times$~0.66~MHz and $\gamma_S$ is determined by the time-of-flight of atoms across the interaction region.  In the bi-chromatic scheme these resonances are broadened approximately by  $(\nu_{2}/\nu_{1}-1)\Delta_{D}\approx$~2.5~MHz due to a mismatch in Doppler shifts arising from the frequency difference of the two optical fields \cite{Gro95}.

If $(\nu_{1}-\nu_{0})\leq\Delta_{D}$ then the step-wise excitation process through the intermediate 5$P_{3/2}$ level contributes to the excitation of the 5$D_{5/2}$ level. This process results in larger linewidth because $\gamma_{P} \gg \gamma_{D}$, where $\gamma_{P}$ is the natural linewidth of the intermediate 5$P_{3/2}$ levels. Thus, the width of the Doppler-free fluorescence resonances depends on a ratio between two processes. Theoretical modeling of the corresponding excitation in Cs atoms \cite{Kargo05} shows that the contribution of the two-photon process prevails at high power. The typical FWHM of experimentally observed fluorescence resonances (Fig. \ref{Figure3}) is approximately 6 MHz, that lies between the two-photon and step-wise limits, taking into account that  power broadening is stronger for the step-wise process. Laser linewidth and magnetic field broadening in the unshielded Rb cell also contribute to the observed width. The shape of the peaks is slightly different to the Lorentzian profile that suggests a presence of distinctive broadening mechanisms.

\subsection{Blue fluorescence dependence on laser detuning }

We use a double frequency sweeping technique to explore the fluorescence amplitude dependence on frequency detuning from the intermediate $5P_{3/2}$ level. This technique allows quick data acquisition that eases requirements for stability of major experimental parameters. The 780 nm laser is slowly  scanned with sweep frequency $f_1$ across a few GHz wide region, while the other laser is swept much faster, with sweep frequency $f_2$ ($f_2\gg f_1$), relative to the position where the sum optical frequency ($\nu_1 +\nu_2$) equals to the frequencies of the transitions from the ground state to the $5D_{5/2}$ level.
Thus, the two-photon excitation condition is met $f_2/f_1$ times during the 780 nm laser scan. This results in appearing a set of fluorescence profiles shifted in frequency of the 780 nm laser. The TDS3012 digital oscilloscope  used for data acquisition is capable to record the whole set of traces.

\begin{figure}
\begin{center}
\includegraphics[angle=0, width=6.5cm]{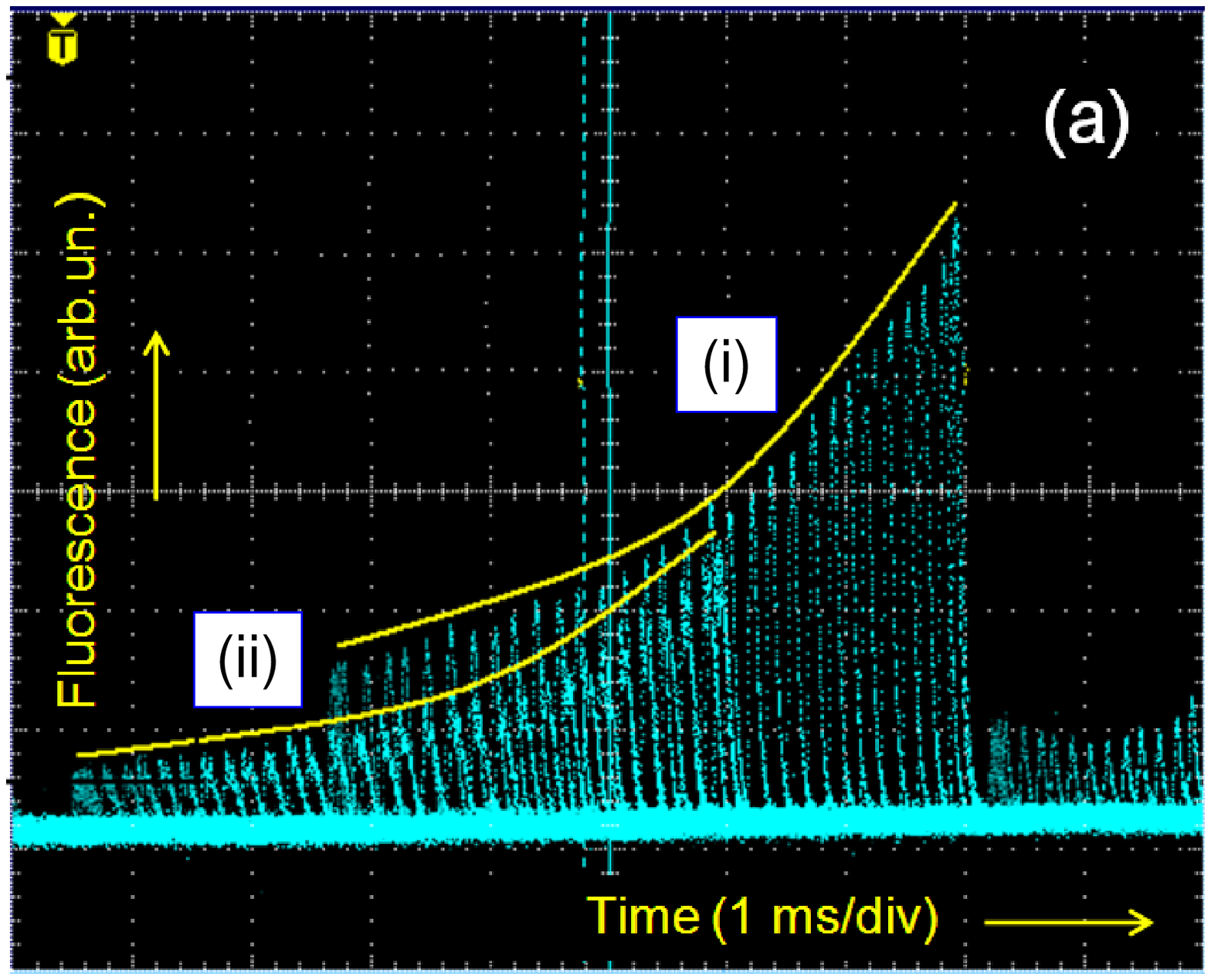}
\includegraphics[angle=0, width=8cm]{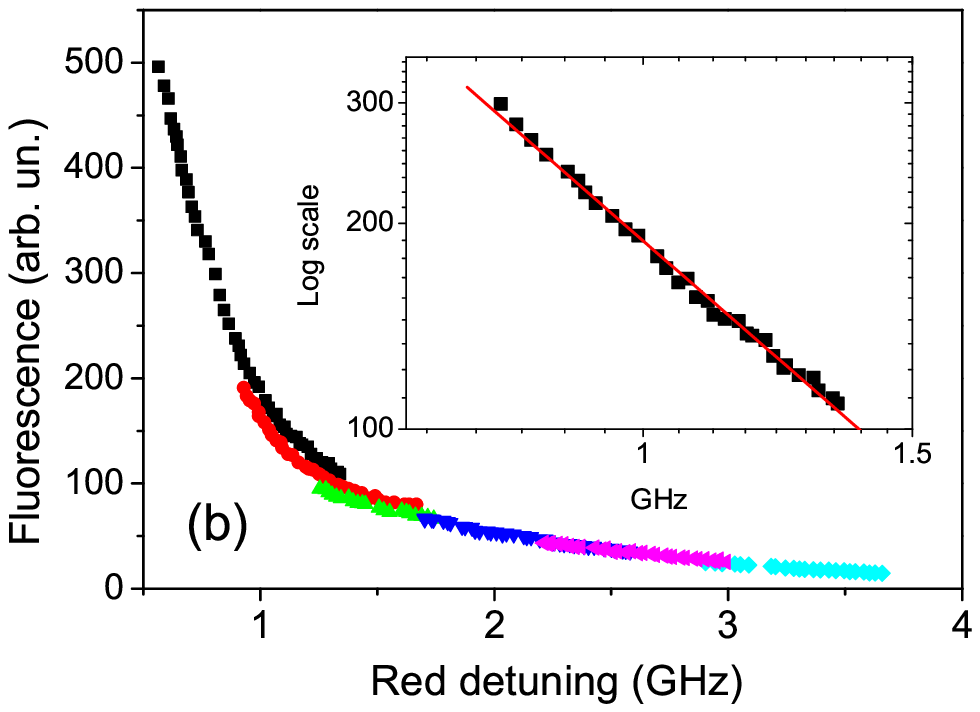}
\end{center}
\caption{a) Curves (i) and (ii) show the evolution of the maximum fluorescence signal obtained on the $5S_{1/2}(F=3)\rightarrow 5D_{5/2}$ and $5S_{1/2}(F=2)\rightarrow 5D_{5/2}$ transitions in $^{85}$Rb and $^{87}$Rb, respectively. Slow 0.5 Hz scan of the 780~nm laser frequency in across the 5$S_{1/2}(F=2)$~$\rightarrow$~5$P_{3/2}$ transitions in $^{87}$Rb is superimposed on fast 50~Hz sweep of the 776~nm laser.  The yellow lines are visual representation of an evolution of peak fluorescence. b) Peak blue fluorescence of $^{85}$Rb atoms excited from the $5S_{1/2}(F=3)$ level to the $5D_{5/2}(F"=5)$ level as a function of the 780 nm laser detuning to lower frequencies from the $5S_{1/2}(F=3)\rightarrow 5P_{3/2}(F'=4)$ transitions. Insert presents the fluorescence amplitude vs. detuning in double-logarithmic scales. \label{Figure4}}
\end{figure}

As an example, Figure \ref{Figure4}a shows the peak fluorescence evolution obtained by the double sweeping technique. The excitation of $^{87}$Rb atoms from the 5$S_{1/2}(F=2)$ level occurs in the spectral region I, which is red-shifted from the Doppler broadened absorption line observed in the axillary cell on the 5$S_{1/2}(F=2)~\rightarrow~5P_{3/2}$ transitions in $^{87}$Rb.
The excitation of $^{85}$Rb atoms from the 5$S_{1/2}(F=3)$ level takes place in the spectral region II, which spectrally coincides with the blue-frequency slope of the same $^{87}$Rb absorption line.  The spectral position and the width of the regions shows that the 776 nm laser is blue-detuned  with respect to the $5P_{3/2}~\rightarrow~5D_{5/2}$ transitions and that the detuning is swept from 0.5 to 1.35 GHz.

The systematic measurements of fluorescence dependence on laser detunings are performed on the $5S_{1/2}(F=3)\rightarrow 5D_{5/2}$ transitions in $^{85}$Rb. The vapour cell temperature is chosen to be 63~$^o$C ($N\simeq4\times 10^{11}$~cm$^{-3}$). As the laser beams are focussed inside the cell, the maximum light intensities at 780~nm and 776~nm are 6.5~W/cm$^2$ and 0.9~W/cm$^2$, respectively. These values are much higher than the saturation intensity for the both one-photon transitions.

Linear fit of experimental points plotted in the log-log scale and shown in insert of Fig.~\ref{Figure4}b  reveals that the slope is equal to -1.98. This means that the fluorescence amplitude varies inversely proportional to the square of the frequency detuning in this spectral region. Despite strong saturation of the single-photon transitions in our experiment, the dependence is in a good agreement with well established theory that describes the spectral lineshape of bi-chromatic two-photon excitation in the low saturation limit \cite{Bjo76,Gro95}, which predicts $(\nu_{1}-\nu_{0})^{-2}$ dependence for the excitation rate. At larger frequency detuning the fluorescence amplitude decays slower, as $(\nu_{1}-\nu_{0})^{-\alpha}$, where $\alpha\simeq1.6$ .

\subsection{Atomic density dependence and detection sensitivity}

\begin{figure}
\begin{center}
\includegraphics[angle=0, width=8cm]{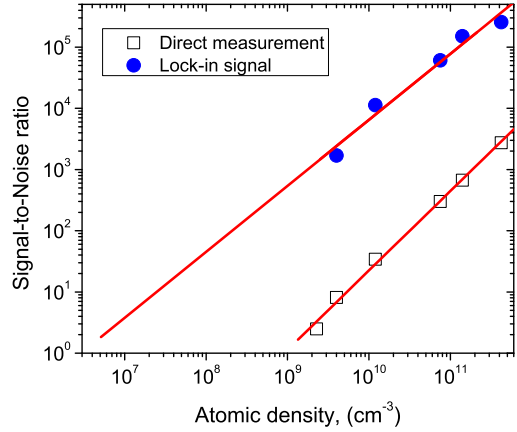}
\end{center}
\caption{ The amplitude of the blue fluorescence resonance of $^{87}$Rb atoms on the $5S_{1/2}(F=2)\rightarrow 5D_{5/2}(F"=4)$ transition as a function of atomic density in the vapour cell. Squares represent the direct measurement of the amplitude of the PMT signal, while circles  show the lock-in amplified signal (time constant is 100 ms) when amplitude modulation at 700 Hz is applied.
\label{Figure5}}
\end{figure}

The atomic density dependence and sensitivity of the method for atom detection are investigated by using signals from the PMT, which are directly recorded or preliminary selectively amplified by a lock-in amplifier. In the latter case either amplitude or frequency modulation of the laser radiation at 780 nm are applied. A mechanical chopper provides 100\% amplitude modulation of the laser beam at 700 Hz, while for the frequency modulation a voltage at 10 kHz is applied to the laser cavity PZT.

To ensure predominantly two-photon nature of the 5$D_{5/2}$ level excitation, the 780 nm laser is red detuned from the strongest 5$S_{1/2}(F=2)~\rightarrow~5P_{3/2}(F'=3)$ transition in $^{87}$Rb atoms.  To keep the laser frequency detuning constant during a rather long period of time, required for changing the atomic density from $2\times 10^9$ to $7\times 10^{11}$~cm$^{-3}$, the following procedure has been used. One transmission peak of the FPC1 cavity is tuned to coincide with the Doppler-free absorption resonance on the 5$S_{1/2}(F=2)\rightarrow 5P_{3/2}(F'=3)$ transition obtained in the auxiliary cell, and then the laser frequency is locked to the next transmission FPC1 peak, which is red-detuned on a free spectral range of the cavity. The position of the reference resonance is checked from time to time insuring that the 780 nm laser frequency is kept detuned on approximately 710 MHz, which is FSR of the FP cavity, from  the 5$S_{1/2}(F=2)~\rightarrow~5P_{3/2}(F'=3)$ transition. The 776 nm laser is scanned across the $5P_{3/2}\rightarrow 5D_{5/2}$ transitions.

Using direct detection we are able to detect the blue fluorescence signal with the signal-to-noise ratio approximately 1 at vapour temperature as low as 10 $^{0}$C at which the atomic density of saturated Rb vapour is $2\times 10^9$~cm$^{-3}$. The signal-to-noise ratio is estimated using an averaged amplitude of the fluorescence signal normalized on the RMS voltage observed for off-resonant laser frequencies.

Figure \ref{Figure5} demonstrates the derived atomic density dependence of the detected fluorescence when intensities of the 780 nm and 776 nm beams in the center of the cell are equal approximately 180 mW/cm$^2$ and 300 mW/cm$^2$, respectively. As expected, the fluorescence signal grows steadily with atomic density. Linear fit to experimental data presented in double-logarithmic scale reveals that the directly observed fluorescence signal is practically proportional to atomic density, while the lock-in processed signal grows slightly faster.

The amplitude and frequency noise of laser light and frequency noise of the detection system limit the detection capability of the method. Signal averaging can increase the signal-to-noise ratio; however, a number of averaging is limited by frequency stability of the both lasers. Taking into account the FWHM of fluorescence resonances ($\simeq$6 MHz) and the typical drift of the FP cavity reference resonance due to temperature and pressure instability (0.4 MHz/sec), we can conclude that averaging the signal over 32 periods of 50 Hz scans of the 776 nm laser  is tolerable because during this time interval (0.64 sec) the drift of the 780 nm laser stabilized over the FP cavity is less than 5\% of the FWHM of the fluorescence resonance. Thus, averaging results in more than 10-fold reduction of the noise uncorrelated with laser scanning and at the same time does not significantly distort the width of recorded spectra.

In order to be in linear regime of the PMT for the whole range of atomic densities, blue fluorescence is recorded at the lowest available PMT voltage. Higher PMT voltage within certain limits can also improve the signal-to-noise ratio. At least 200-fold increase of the signal under the same noise level can be obtained in our experimental conditions. Altogether with averaging the signal this could allow us to detect atomic density lower than $2\times10^6$~cm$^{-3}$. An estimation for the volume of the interaction zone, which in our case is 0.35 cm$^{-3}$, leads to a conclusion that a detection threshold is approximately $7\times10^5 $ atoms.

An improvement of the detection capability can be accessed by applying an amplitude modulation and a lock-in amplification.
Our estimations based on measured spectra of the amplitude noise of fluorescence signals show that signal-to-noise ratio can be increased by a factor of 750 and 2000 for the integration time of 100 ms and the modulation frequency of 700 Hz and 10 kHz, respectively.  Ultimately the detection threshold is reduced to approximately  1000 Rb atoms in the interaction volume. This result can be further improved primarily by better collection of emitted blue fluorescence (in our case the collection efficiency is approximately 1.8\%) and longer, up to minutes, signal averaging accessible by better laser stability.
Optical pumping through the $D_{1}$ line could increase a number of atoms on a resonant sublevel and therefore increase the fluorescence signal intensity.


\section{Sum frequency locking of two independent lasers}

As Doppler-free fluorescence resonances occur when the sum frequency of the two lasers ($\nu_1$ + $\nu_2$) is equal to frequencies of the $5S_{1/2}\rightarrow 5D_{5/2}$ transitions, they can be used as a reference for sum frequency stabilization using a single servo system. The similar idea has been  implemented with two independent lasers locked to a narrow CPT-type absorption resonance ($\Lambda$-resonance), which is also has the two-photon origin, however, in that case the locking results in differential frequency stabilization \cite{Aku91}.

The sum-frequency locking of two lasers can be implemented  using a standard locking technique based on the frequency modulation and the lock-in amplification  and one of the sub-Doppler fluorescence resonances shown in Fig. \ref{Figure3}.  It  makes no difference whether the feedback applied to the 780 nm or 776 nm lasers and which laser is modulated. The sum frequency ($\nu_1$ + $\nu_2$) can be stabilized even if one laser is free running. Once the lock is engaged the servo system compensates random frequency excursions of the free-running laser by changing correspondingly the frequency of the stabilized laser to keep the sum frequency fixed.


To demonstrate the direct sum frequency stabilization using one servo system, the 780 nm laser is approximately 600 MHz blue detuned  from the $5S_{1/2}(F=2)\rightarrow 5P_{3/2}(F'=3)$ transition in $^{87}$Rb, while frequency of the 776 nm laser is tuned in a such way that an excitation of $^{87}$Rb atoms to the $5D_{5/2}(F"=4)$ level occurs.

Using a PID servo system which provides only slow feedback (from DC to ~1~kHz) to the piezo-electric transducer that controls the 780 nm laser cavity, the sum frequency instability  estimated from the error signal approximately equals to 240 kHz. This value corresponds to the relative  instability $\delta\nu/(\nu_1$ + $\nu_2)$ of $3.6\times10^{-10}$.  The sum frequency instability can be essentially reduced by a proper optimization of the reference line and by locking of the free-running laser to a tunable FP cavity. The suggested and demonstrated sum frequency stabilization could be useful for Rydberg atoms excitation experiments \cite{SW} and in intervening spectral regions devoid of strong atomic transitions.
As the blue fluorescence signal depends on frequency detuning, this can then be used to derive an absolute frequency of the 780~nm laser in the vicinity of the Rb $D_2$ line.  It is envisaged that this could form a basis for an additional level locking scheme based on the amplitude of the fluorescence resonance.

\subsection{Polarization spectroscopy of the 5$P_{3/2}\rightarrow5D_{5/2}$ transition in Rb}

\begin{figure}
\begin{center}
\includegraphics[angle=0, width=8cm]{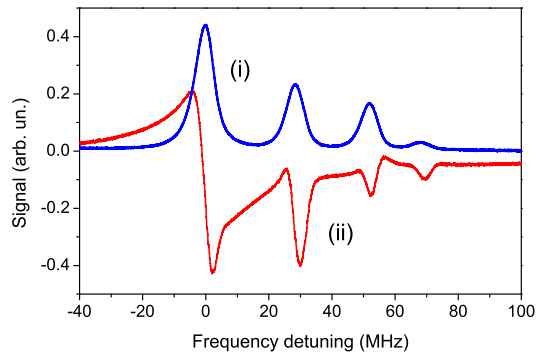}
\end{center}
\caption{Polarization spectroscopy of the 5$S_{1/2}(F=2)\rightarrow5D_{5/2}$ transition in $^{87}$Rb. Curves (i) and (ii) represent blue fluorescence and the intensity of light at 776~nm transmitted through the cell and the almost crossed polarize, respectively, as a function of the frequency detuning of the 776~nm laser.}
\label{Figure6}
\end{figure}

Even small frequency modulation applied for frequency locking results in additional broadening of the laser linewidth, which is often undesirable. A dispersive-shaped polarization resonance is an ideal reference line for the modulation-free frequency locking.

The polarization rotation of resonant monochromatic laser light that excites the two-photon transition was demonstrated in \cite{Hamid03}, and we implement a polarization spectroscopy set-up for the bi-chromatic excitation scheme. Considering that independent setting of polarizations of  laser beams is important for maximizing the polarization signals, the bi-chromatic excitation scheme allowing this arrangement possesses an obvious advantage compared to the monochromatic excitation scheme.

In the present experiment, by an analog to the conventional polarization spectroscopy of one-photon transitions \cite{Dem98}, the 776~nm laser beam analyzes birefringence of the Rb vapour induced by the circularly polarized 780~nm laser in the vicinity of the $5P_{3/2}\rightarrow 5D_{5/2}$ transition. This  involves only minor changes to the optical scheme presented in Fig. \ref{Figure1}b. The intensity of the linearly polarized 776~nm laser beam  transmitted through the vapour cell and a polarizer is detected by  a photodiode. When the polarizer is almost crossed the spectral dependance of the transmitted probe-laser intensity reveals resonances with the well-pronounced dispersive shape. To make this signal suitable for sum frequency locking a non-resonant DC background should be subtracted using a differential signal from two photodiodes \cite{Pea02}.  The steep slope of the  polarization signal, shown in Fig. \ref{Figure8}, coincides with the top of fluorescence resonance recorder simultaneously on the strongest $5S_{1/2}(F=2)\rightarrow 5D_{5/2}(F"=4)$ transition for $^{87}$Rb atoms.

\section{Conclusion}

We study two-photon and step-wise bi-chromatic excitations of the 5$D_{5/2}$ level in $^{85}$Rb and $^{87}$Rb atoms by monitoring blue fluorescence emitted by the atoms decaying from the level 6$P_{3/2}$.
The double frequency sweeping technique allows quick data acquisition and relaxes requirements for stability of major experimental parameters.
The dependence of blue fluorescence as a function of atomic density is investigated in the temperature range of 10 to 70 $^o$C.

The fluorescence amplitude varies inversely proportional to the square of the frequency detuning from the intermediate $5P_{3/2}$ level.
Despite strong saturation of the single-photon transitions, the dependence is in a good agreement with the theoretical prediction of the two-photon excitation rate obtained for the low saturation case.

The sensitivity of the bi-chromatic scheme for atom detection using direct fluorescence observation and lock-in amplification has been estimated.
A novel method for the sum-frequency stabilization of two free-running lasers has been implemented using the Doppler-free two-photon excitation and polarization resonances. The estimated value of the relative sum-frequency instability is lower than $4\times10^{-10}$.

\begin{ack}
This project is supported by the Swinburne University of Technology Strategic Initiative grant.
\end{ack}

\end{document}